\documentclass[useAMS,usenatbib]{mnras}
\usepackage{graphicx}
\usepackage{longtable}
\usepackage{graphicx}
\usepackage{hyperref}
\usepackage{amsmath}
\usepackage[singlelinecheck=false]{caption}
\title[IFU spectroscopy of southern PN VI]
  {IFU spectroscopy of southern PN VI: \\
 The extraordinary chemo-dynamics of Hen 2-111}
\author[M. A. Dopita et al.]
  {M.A. Dopita,$^{1}$ A. Ali$^{2,3}$, A. I. Karakas$^{4}$, D. Goldman$^{5}$, M. A. Amer$^{2,3}$ \& R. S. Sutherland$^{1}$ \\
  $^1$Research School of Astronomy and Astrophysics, Australian National University, Cotter Rd., Weston ACT 2611, Australia \\
  $^2$Astronomy Dept, Faculty of Science, King Abdulaziz University, Jeddah, Saudi Arabia \\
  $^3$Department of Astronomy, Faculty of Science, Cairo University, Egypt \\
  $^4$Monash Centre for Astrophysics, School of Physics and Astronomy, Monash University, Victoria 3800, Australia \\
  $^5$Astrodon Imaging, Roseville, CA 95661, USA \\
  }
\date{Released 2016 Xxxxx XX}

\pagerange{\pageref{firstpage}--\pageref{lastpage}} \pubyear{2017}

\def\LaTeX{L\kern-.36em\raise.3ex\hbox{a}\kern-.15em
    T\kern-.1667em\lower.7ex\hbox{E}\kern-.125emX}

\begin{document}

\label{firstpage}

\maketitle

\begin{abstract}
In this paper we present integral field spectroscopy of the extraordinary Type I bipolar planetary nebula Hen 2-111. In the lobes we map fast moving knots of material with [N II]$\lambda 6584$/H$\alpha$ ratios up to 12, and with radial velocities relative to systemic from  -340\,km/s up to +390\,km/s. We find evidence of a bipolar ejection event at a velocity $\sim 600$\,km/s from the central star (assumed to be a binary), which occurred about 8000\,yr ago. The fast moving material is chemically quite distinct from the lower velocity gas in the bipolar lobes., and displays very high N abundances. We show that the fast moving N-rich knots are not photoionised by the central star, and have constructed detailed shock models for the brightest knot. We find a pre-shock density $\sim 6$\,cm$^{-3}$, and a shock velocity $\sim150$\,km/s.The shock is not fully radiative, being only $\sim 600$\,yr old. This shocked gas is partially H-burnt, with a helium abundance by mass exceeding that of hydrogen, and is interacting with partially H-burnt material ejected in an earlier episode of mass loss. We conclude that the high-velocity material and the bipolar shell must have originated during the late stages of evolution of a common-envelope phase in a close binary system.
\end{abstract}


\section{Introduction}

The Type I planetary nebula (PN) Hen 2-111 (PN G315.0-00.3) first discovered by  \citet{Henize67}, presents one of the more extreme examples of nebular N- and He- enrichment in the southern sky \citep{Perinotto98, Pottasch00, Marigo03}. 
\citet{Webster78} found that the central core of the nebula is surrounded by an extensive bipolar halo with its long axis at about PA=130$^0$  and covering about 10 arc min. on the sky in this direction. She showed that this halo is characterised by a range of radial velocities of -150 to +275 km/s with respect to the systemic velocity of the core. Eleven years later \citet{Meaburn89} subjected the nebula to much higher resolution \'echelle spectrometric observations, discovering gas out to a maximum (Heliocentric) radial velocity of +360\,km/s  in the western lobe, and -380\,km/s in the eastern extremity.

The explanation of such high velocities has remained somewhat of a mystery. The idea that a supernova \citep{Webster78} or else a nova or a similar eruptive event \citep{Meaburn89} has been advanced to explain the high velocities. We should note that Hen 2-111 is one of three PN which show extraordinary dynamics.  \citet{Lopez93} found that the planetary nebula Fleming 1 has  a symmetric string of knots emerging from the central nebula, which reach a radial velocity of $\pm74$\,km/s at the extremities. These were interpreted in terms of a bipolar rotating episodic jet (BRET) model.  In MyCn 18 (PN G307.5-04.9) \citet{Bryce97} found hypersonic knots with velocities $\sim 500$\,km/s situated along the polar axis of this famous Hourglass Nebula. These were also interpreted on the basis of the BRET model.

The theoretical justification for the BRET model was developed by \citet{Soker92}. In this, a star in a common envelope binary system blows a collimated wind during the evolution from the Asymptotic Giant Branch (AGB )phase to become the central star of a PN. Two kinds of jet can be distinguished depending on the orbital separation of the central binary. For larger orbital separations, $a \geqslant 5~R_{\odot}$, slow, heavy jets are formed with velocities $\sim 50$\,km/s and total mass $\geqslant 10^{-3}~M_{\odot}$ while for smaller orbital separations, fast-light jets (which can continue to be active after the system turns into a PN) are formed with velocities up to $300$\,km/s and total mass $\sim 10^{-4}~M_{\odot}$.

In order to better understand the case of Hen 2-111, we undertook an integral field survey of selected regions in the core and in the bipolar lobes of this very spatially extended nebula. The considerable advantages of using integral field unit (IFU) spectroscopy as opposed to long slit spectroscopy  were outlined in an earlier paper by \citet{Ali16}.

\begin{figure*}
  \includegraphics[scale=0.7]{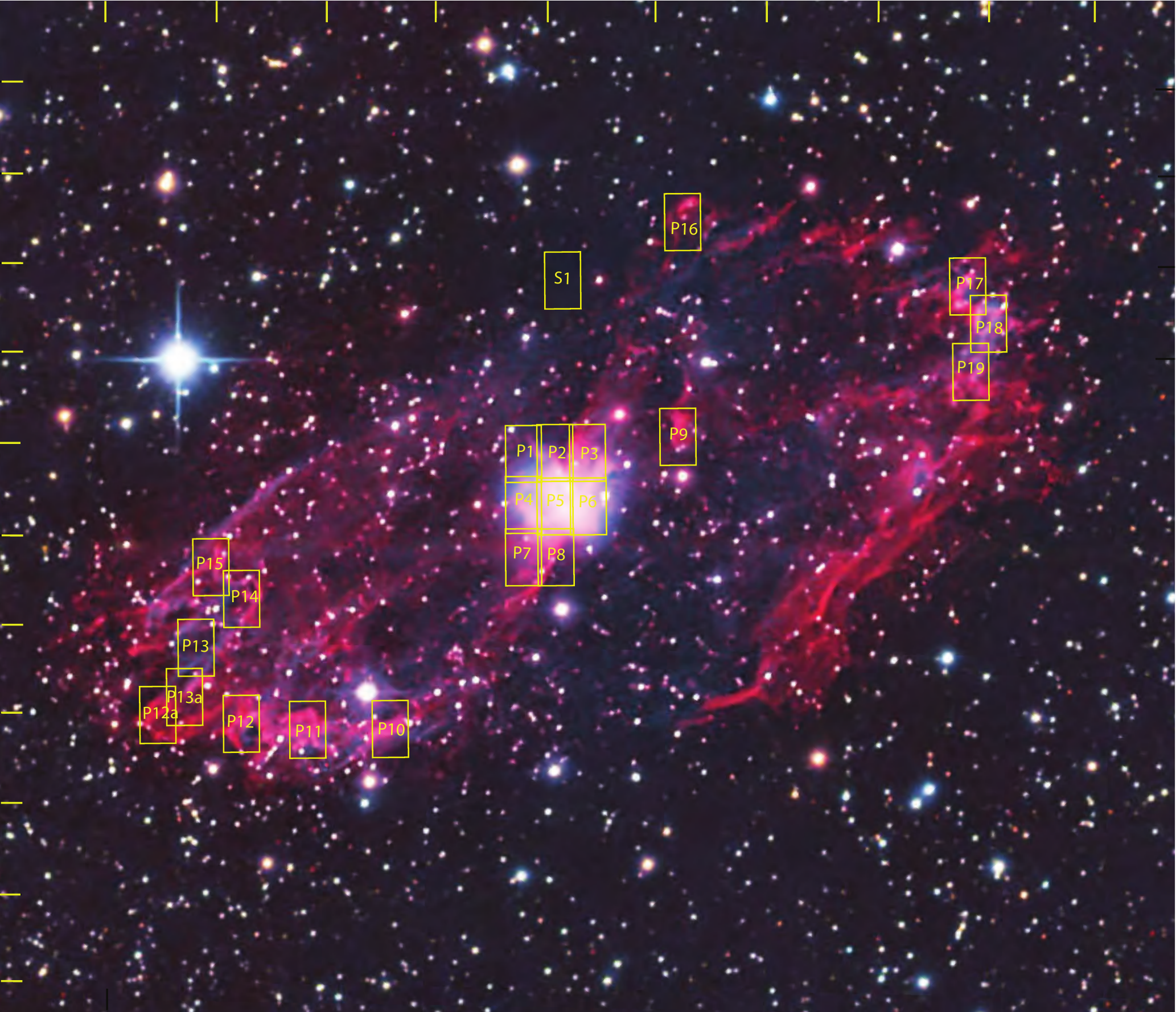}
  \caption{The positions observed with WiFeS plotted on an H$\alpha$ + [O III] + stellar continuum image of Hen 2-111. This image is available at:  {\url {astrodonimaging.com/gallery/henize-2-111/}}} \label{fig1}
 \end{figure*}

This paper is the sixth in a series presenting results of integral field spectroscopy on southern planetary nebulae using the Wide Field Spectrograph (WiFeS) instrument \citep{Dopita07,Dopita10}.  The earlier papers have examined the large, evolved and interacting planetary nebula PNG 342.0-01.7 \citep{Ali15b}, provided a detailed analysis of four highly excited non-type I PN which casts doubt on the general applicability of the WELS classification of planetary nebula nuclei \citep{Basurah16}, presented maps and abundance analysis of four PN: M3-4; M3-6; Hen 2-29; Hen 2-37  \citep{Ali16}, provided a detailed analyse of the low-excitation young PN IC 418 \citep{Dopita17a}, and presented spectroscopy and morphology of four southern Galactic planetary nebulae Hen\,2-141, NGC\,5307, IC\,2553, and PB\,6  \citep{Ali17}.

The distance to Hen 2-111, as for many individual galactic PN, remains somewhat uncertain. However \citet{Frew08} obtains good agreement between the kinematical method ($2100\pm500$\,pc) and the extinction method ($2200\pm500$\,pc). Further confidence in these estimates is obtained using the statistical methods of distance determination. \citet{Frew16} obtains $2090\pm700$\,pc based on the E(B-V) and a radius of 21.75arc sec. Finally using the 5cm radio flux, \citet{Ali15a} derives a distance of 2160\,pc. For the purposes of this paper we adopt a distance of 2100\,pc with a probable error of 500\,pc.

The current paper is organised as follows. In Section \ref{Obs} we present the observations and data reduction technique, while the results are given in terms of the imaging of central core and the chemo-dynamics of the bipolar halo in Section \ref{results}. These results are discussed in Section \ref{discussion}. In the final section, we summarise our results, and examine these in the context of the theory of the late evolution of AGB stars.

\section{Observations \& data reduction}\label{Obs}

\begin{table*}
 \centering
 \small
   \caption{The log of  WiFeS observations of Hen 2-111}
    \label{Table1}
   \scalebox{0.95}{
  \begin{tabular}{lcccc}
 \hline
   Position & RA  & Dec & Date  & Exp. Time \\
& (J2000) &  (J2000) &   &   (s)  \\
   \hline \hline
Ref. Star &  14:33:18.60 &  -60:50:50.7 & & \\
Sky Ref.  &  14:33:18.70 &  -60:47:14.0 & & \\
P18 &  14:32:40.20 &  -60:47:40.0 &  22 April 2017 & $2\times900$ \\
P19 &  14:32:41.50 &  -60:48:10.0  &  22 April 2017 & $2\times900$ \\
P17 &  14:32:42.20 &  -60:47:15.0  &  22 April 2017 & $2\times900$ \\
P16 &  14:33:08.00  &  -60:46:31.0 & 30 May 2017 & $2\times900$ \\
P9  &  14:33:08.20   &  -60:48:55.0  & 22 April 2017& $3\times180$ \\
P3  &  14:33:16.13  & -60:49:06.5  & 21 April 2017 & $3\times180$ \\
P6  &  14:33:16.13  & -60:49:40.6  &  22 April 2017 & $3\times180$ \\
P2  &  14:33:19.04  & -60:49:06.5  & 21 April 2017 & $3\times180$ \\
P5  &  14:33:19.04  & -60:49:40.6  & 22 April 2017 & $3\times180$ \\
P8  &  14:33:19.04  & -60:50:15.0  &  22 April 2017 & $3\times180$ \\
P1  &  14:33:21.89  & -60:49:06.5  & 21 April 2017 & $3\times180$ \\
P4  &  14:33:21.89  & -60:49:40.6  & 21 April 2017 & $3\times180$ \\
P7  &  14:33:21.89  & -60:50:15.0  & 22 April 2017& $3\times180$ \\
P10 &  14:33:34.20 & -60:52:12.0  & 30 May 2017 & $2\times900$ \\
P11 &  14:33:41.60 & -60:52:07.0  & 23 April 2017 & $2\times900$ \\
P12 &  14:33:47.40 & -60:52:07.0  &  23 April 2017 & $2\times900$ \\
P14 &  14:33:47.40 & -60:50:45.0 &  22 April 2017  & $2\times900$ \\
P15 &  14:33:50.40 & -60:50:23.0  &  23 April 2017 & $2\times900$ \\
P13 &  14:33:51.50 & -60:51:17.0  &  23 April 2017 & $2\times900$ \\
P13a & 14:33:52.90 & -60:51:48.0  & 29 May 2017 & $2\times900$ \\
P12a & 14:33:55.00 & -60:52:00.0  &  23 April 2017 & $2\times900$ \\
 \hline
 \end{tabular}}
\end{table*}

The integral field spectra of He2-111 were obtained between 21 April 2017 and 30 May 2017 using the WiFeS instrument \citep{Dopita07,Dopita10} mounted on the 2.3-m ANU telescope at Siding Spring Observatory. This instrument delivers a field of view of 25\arcsec $\times$ 38\arcsec at a spatial resolution of either 1.0\arcsec $\times$ 0.5\arcsec or 1.0\arcsec $\times$ 1.0\arcsec, depending on the binning on the CCD. In these observations, we operated in the binned 1.0\arcsec x 1.0\arcsec mode. The data were obtained in the high resolution mode $R \sim 7000$ (FWHM of $\sim 45$ km/s) using  the B7000 \& R7000 gratings in each arm of the spectrograph, with the dichroic cut set at 560nm (using the RT560 dichroic). For details on the various instrument observing modes, see \citet{Dopita07}.

The log of the observations is given in Table \ref{Table1}. The corresponding positions are shown in Figure \ref{fig1} on the colour image made by Don Goldman. This image is a composite of 8.5 hrs  exposures using 3 nm H$\alpha$ filter, 5.5 hrs using a  3 nm [O III] filter and 1 hour of RGB imaging for star colors.Here the H$\alpha$ is colour-mapped to red/magenta and [O III] mapped to turquoise to create a ``natural color" image. RGB star colours were created based upon a G2V calibration. This is the first image processed from the new iTelescope.net site at Siding Springs in Australia with a new PlaneWave CDK20 telescope mounted on a PlaneWave Ascension A200h mount.

The observations of each region outside the central core of the nebula were interspersed with a single sky exposure of 900s at our sky reference position, S1 in the figure, to permit accurate subtraction of the sky background. The nights of 21 April, 23 April and 30 May were non-photometric, resulting in somewhat lowered signal to noise, and greater uncertainty in the flux calibration. However, the seeing remained very stable during all of the observing, ranging from 1.0 - 1.4 arc sec.

The wavelength scale was calibrated using the Ne-Ar arc Lamp throughout the night. Arc exposure times are 100s at B7000 and 9s for the  R7000 grating. Flux calibration was performed using the STIS spectrophotometric standard stars HD\, 111980 \& HD\,160617 \footnote{Available at : \newline {\url{www.mso.anu.edu.au/~bessell/FTP/Bohlin2013/GO12813.html}}}. In addition, a B-type telluric standard HIP\,45754 was observed to better correct for the OH and H$_2$O telluric absorption features in the red. The separation of these features by molecular species allows for a more accurate telluric correction by accounting for night to night variations in the column density of these two species. All data cubes were reduced using the PyWiFeS \footnote {\url{http://www.mso.anu.edu.au/pywifes/doku.php.}} data reduction pipeline (\citet{Childress14}). 

Spectra and velocity channel images were extracted using {\tt QFitsView v3.1 rev.741}\footnote{{\tt QFitsView v3.1} is a FITS file viewer using the QT widget library and was developed at the Max Planck Institute for Extraterrestrial Physics by Thomas Ott.}.  The resulting 2-D images were manipulated using the IRAF code \footnote{Available at :  {\url{iraf.noao.edu}}}. We first smoothed the data using a 0.3 arc sec. Gaussian kernel, then trimmed the H$\alpha$ images at a flux level of $2\times10^{-17}$ erg/s/cm$^2$/arcsec$^2$/\AA\ and finally divided the [N II] image by the trimmed H$\alpha$ images to obtain line ratio maps. These were then examined and saved in the form presented here using {\tt QFitsView v3.1 rev.741}.

\section{Results}\label{results}
\subsection{Imaging of the Nebular Core}
\begin{figure*}
  \includegraphics[scale=0.35]{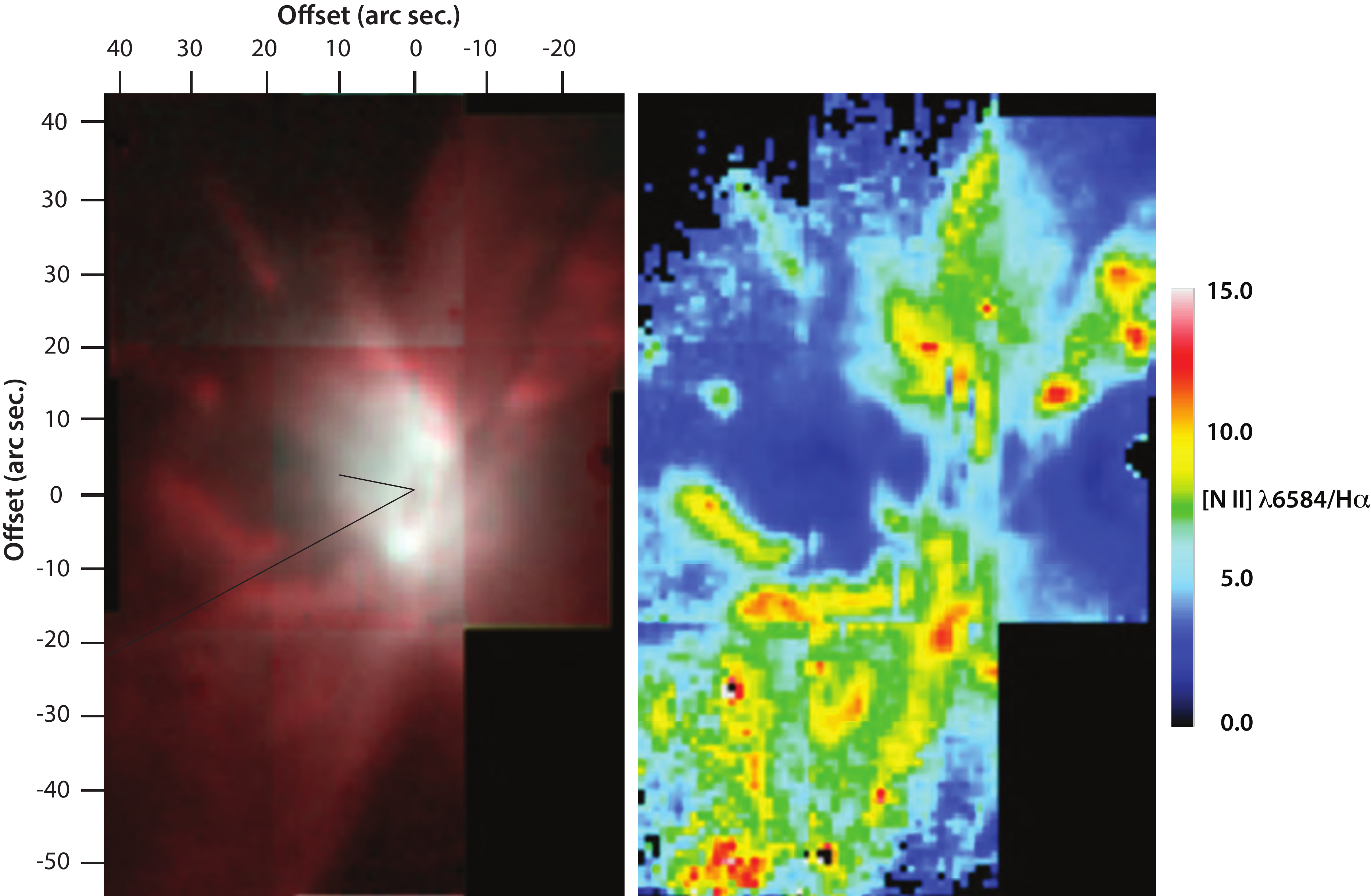}
  \caption{Left: The central regions of Hen 2-111 in red, [N II] and in green and blue, H$\alpha$. A logarithmic stretch has been used to bring out the fainter regions. Right: The measured [N II]$\lambda 6584$/H$\alpha$ ratio. The images have been trimmed at an H$\alpha$  flux level of $2\times10^{-17}$ erg/s/cm$^2$/arcsec$^2$/\AA\ to remove noisy data. Note the misalignment between the axis of the inner ring structure (marked by the short black line), and the axis of the outer bipolar halo (shown by the long black line) from Figure \ref{fig1}.} \label{fig2}
 \end{figure*}

From the eight contiguous positions in the centre of the nebula, we have constructed a mosaic in the emission lines of H$\alpha$ and [N II], which is shown in the left-hand panel of Figure \ref{fig2}. This shows the central ring and the outer almost radial filaments to be relatively bright in [N II], and nicely brings out the overall hourglass shape of the nebula. The central ring is inclined with respect to the symmetry axis.

The right-hand panel of Figure \ref{fig2} shows the measured [N II]$\lambda 6584$/H$\alpha$ ratio across the central core. This ratio reaches a peak value of $\sim 12$, and high values of the ratio are associated with the ionisation front regions along the filamentary structures. Note that the central ring is not particularly strong in [N II]. The two vertical stripes  each one pixel wide which lie just north of the central ring are artefacts due to line saturation on the chip. The dark circular region on the western extremity of the image is caused by improper subtraction of a bright star.

 It is important in the analysis which follows to have some idea of the angle of inclination of the whole structure to the line of sight. We estimate this using the region brightest in H$\alpha$ which appears as the greenish-blue region in the left hand panel Figure \ref{fig2}. This region has dimensions $13\times25$\,arc sec. which would imply that it lies at an angle of $31^o$ to the plane of the sky. This is somewhat less than the $45^o$ estimated by \citet{Meaburn89} on the basis of the observed dynamics of the inner region. For the purpose of this paper we adopt the mean of these two estimates; $38^o$ to the plane of the sky.

 \subsection{Excitation Maps of the Bipolar Halo}
\begin{figure*}
  \includegraphics[scale=0.4]{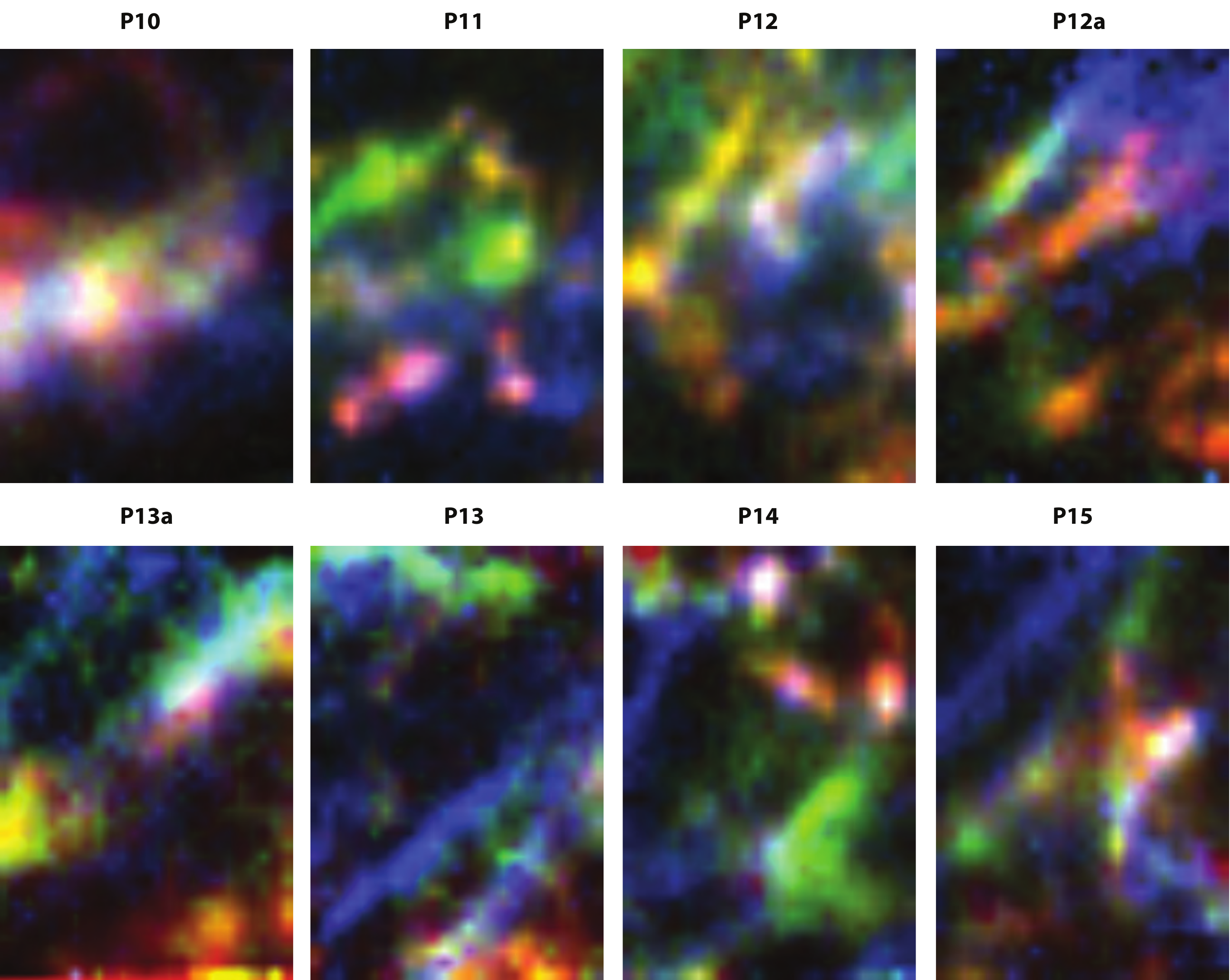}
  \caption{Excitation maps of the fields observed in the eastern lobe of Hen 2-111. These images are linearly scaled in [N II] (red channel), H$\alpha$ (green channel) and [O III] (blue channel).} \label{fig3}
 \end{figure*}

In Figures \ref{fig2} and \ref{fig3} we present excitation maps in [N II] (red), H$\alpha$ (green) and [O III] (blue) of the individual fields observed in the bipolar halo. These images are presented with linear scaling in all channels, scaled from zero flux up to the maximum flux in each of the colour channels, so these images represent relative fluxes, rather than providing ``true colour" images. The [N II] rich knots are clearly visible, as well as the tendency of the [O III] to form linear structures aligned with the axis of the outer bipolar lobes.

\begin{figure*}
  \includegraphics[scale=0.4]{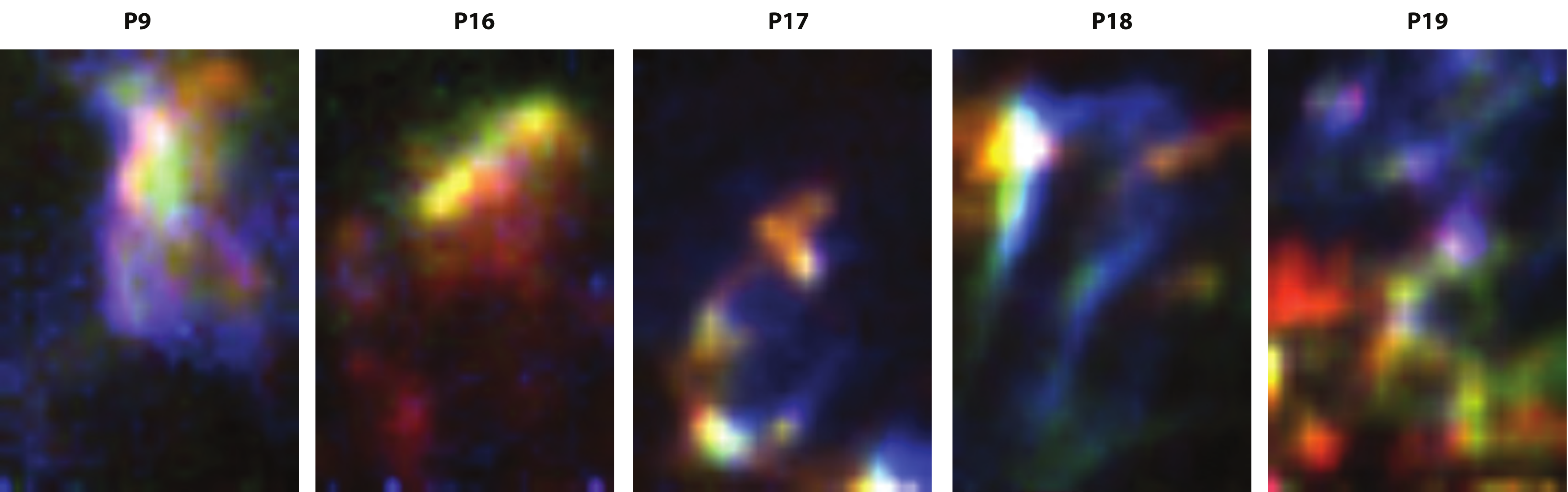}
  \caption{Excitation maps as in  Figure 3 but for fields observed in the western lobe of Hen 2-111.} \label{fig4}
 \end{figure*}

\subsection{Chemo-dynamics of the Bipolar Halo}\label{chemo}

\begin{figure*}
  \includegraphics[scale=0.5]{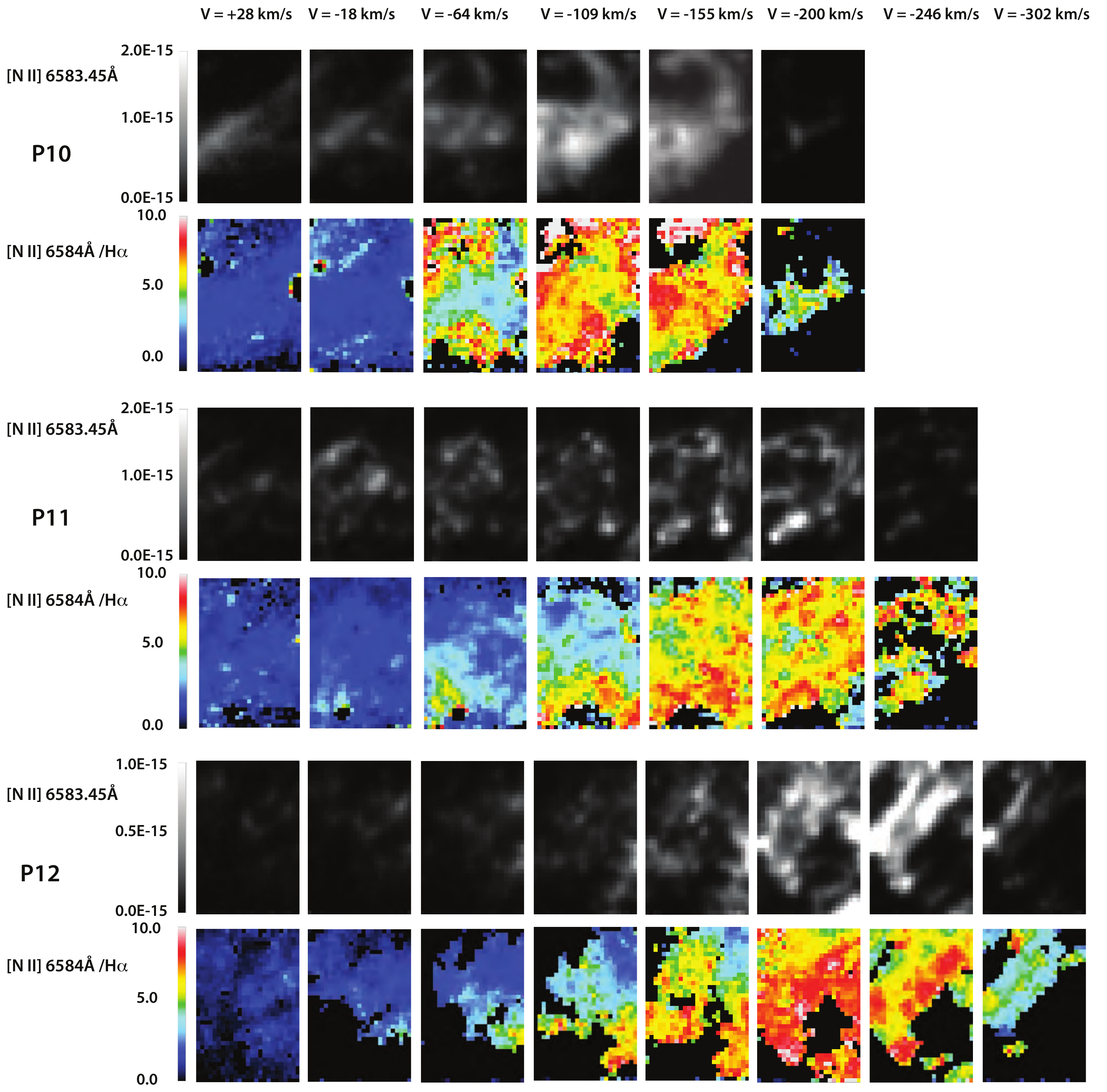}
  \caption{Velocity channel maps of the [N II] emission and the corresponding [N II]$\lambda 6584$/H$\alpha$ ratio for P10, P11 and P12 in the eastern lobe. The scale of the intensity plot is linear and common to all panels. This scale is given on the left hand side of the figure. The [N II]$\lambda 6584$/H$\alpha$ ratio is also linear in scale, and the scale is given on the left hand side of the figure. Velocities for each channel are given relative to the systemic velocity of Hen 2-111. Note in this, and in subsequent figures that the [N II]$\lambda 6584$/H$\alpha$ ratio is much higher in the fast-moving material.} \label{fig5}
 \end{figure*}

\begin{figure*}
  \includegraphics[scale=0.52]{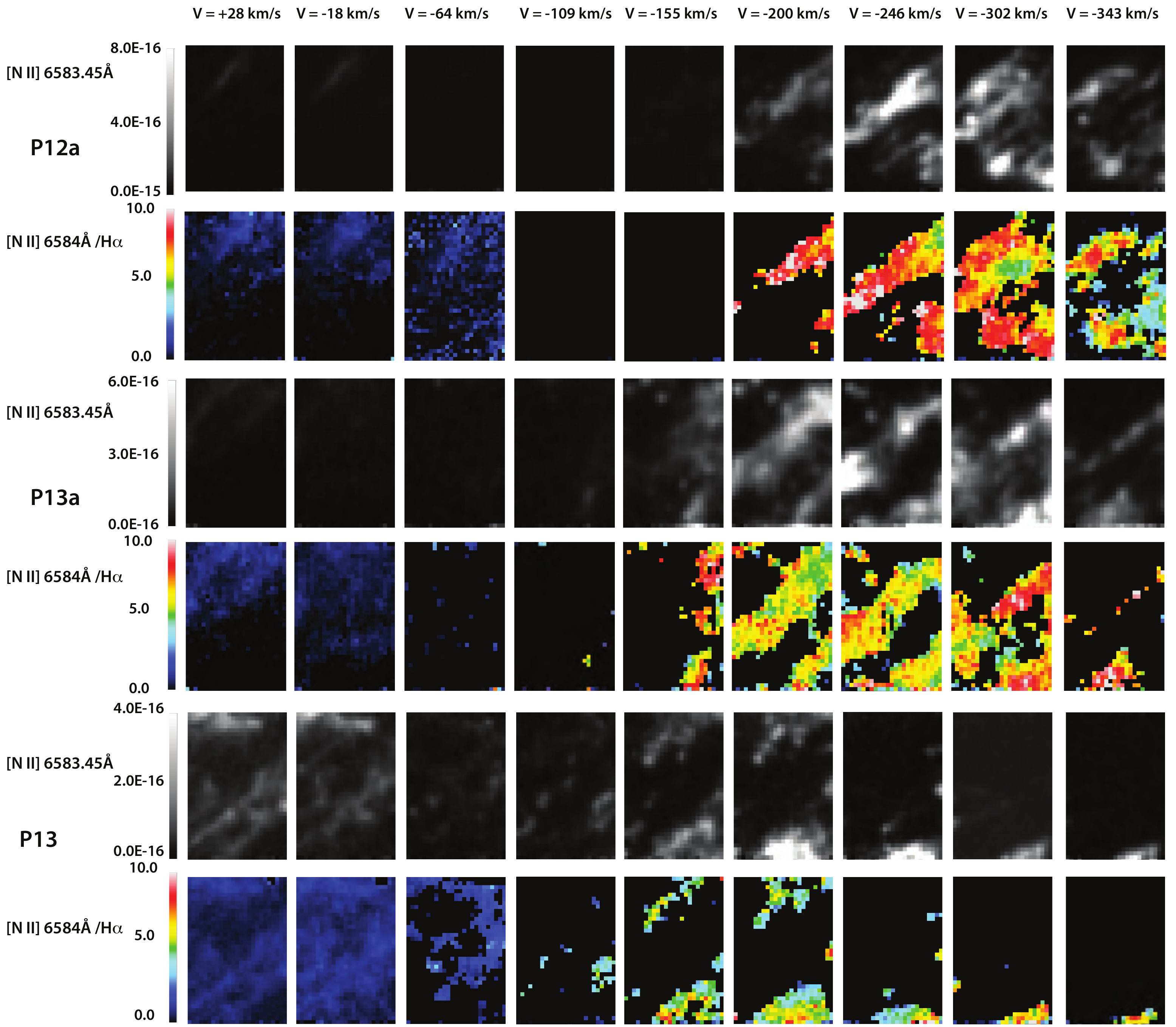}
  \caption{Velocity channel maps of the [N II] emission and [N II]$\lambda 6584$/H$\alpha$ ratio for P12a, P13a and P13  in the eastern lobe.} \label{fig6}
 \end{figure*}
 
 \begin{figure*}
  \includegraphics[scale=0.52]{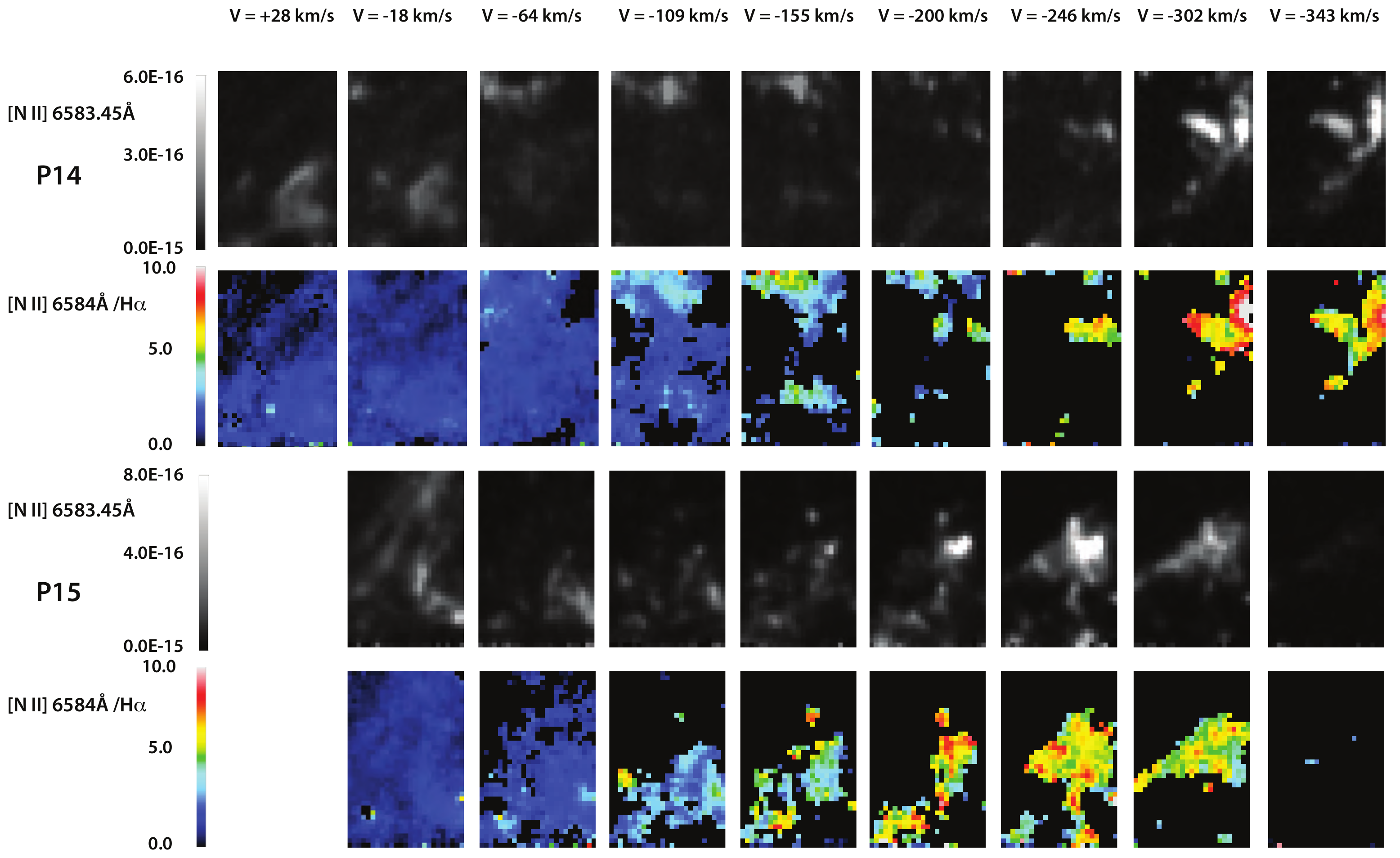}
  \caption{Velocity channel maps of the [N II] emission and [N II]$\lambda 6584$/H$\alpha$ ratio for P14 and P15 in the eastern lobe.} \label{fig7}
 \end{figure*}

 \begin{figure*}
  \includegraphics[scale=0.52]{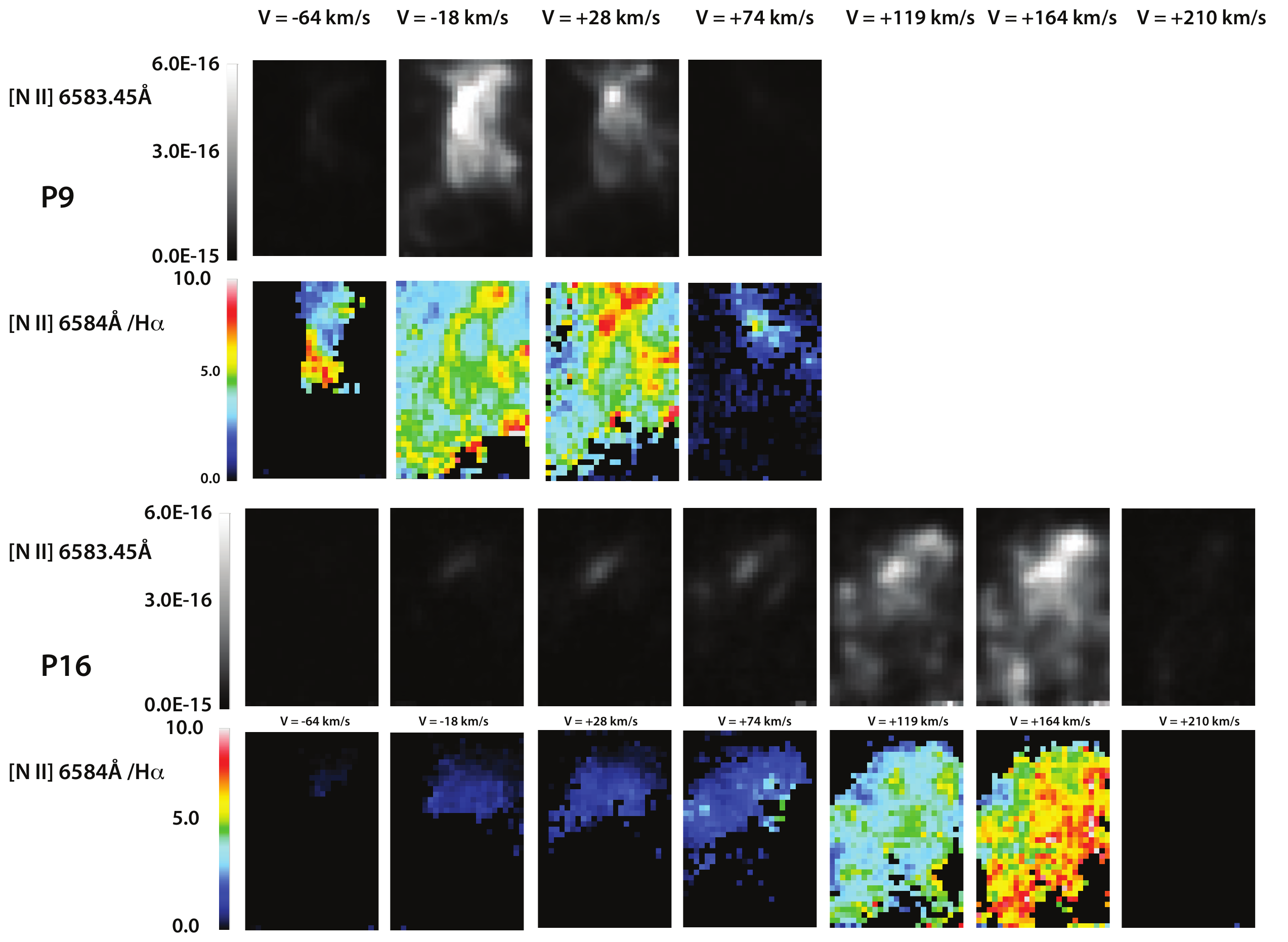}
  \caption{Velocity channel maps of the [N II] emission and [N II]$\lambda 6584$/H$\alpha$ ratio for P9 and P16  in the western lobe.} \label{fig8}
 \end{figure*}

\begin{figure*}
  \includegraphics[scale=0.5]{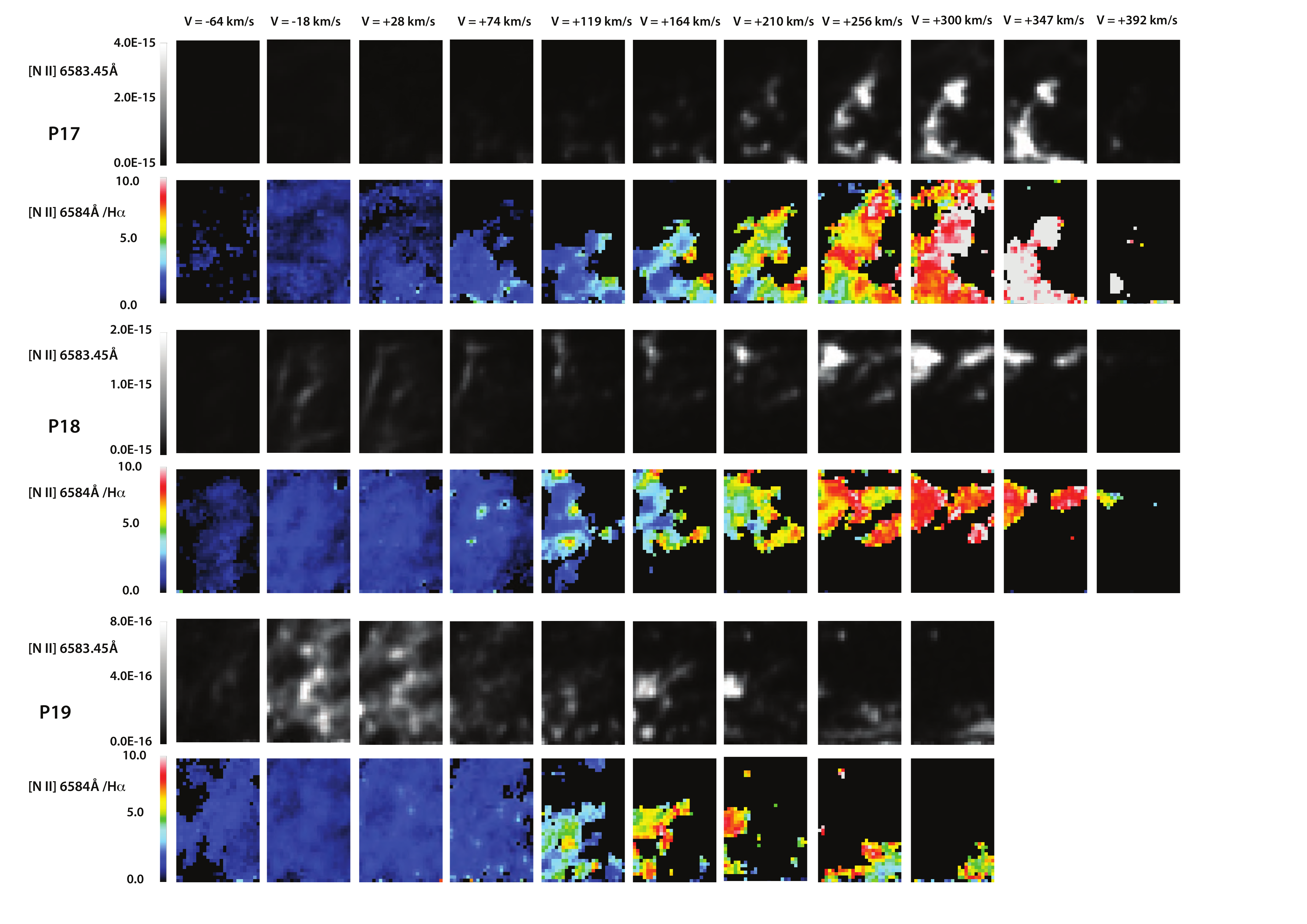}
  \caption{Velocity channel maps of the [N II] emission and [N II]$\lambda 6584$/H$\alpha$ ratio for P17, P18 and P19 in the western lobe. } \label{fig9}
 \end{figure*}

From the data cubes, it is evident that the high-velocity material in Hen 2-111 is characterised by very enhanced  [N\,II] $\lambda 6584$/H$\alpha$ ratios compared with the material at systemic velocity. Such an enhancement was first noted by \citet{Meaburn89}. In order to investigate this effect, we have analysed all of our data cubes taken in the lobes of Hen 2-111 by velocity. Figures \ref{fig3} to \ref{fig7} show the [N II] intensity (on a linear stretch), and the corresponding [N II] $\lambda 6584$/H$\alpha$ ratio map as a function of radial velocity relative to the systemic velocity of the PN. Each velocity channel is separated from the others by a resolution element of the spectrograph ($\sim 45$\,km/s). In these the local Galactic contribution to the emission has been fully corrected for, since each data cube has had the sky contribution from the sky reference position (S1 in Figure \ref{fig1}) subtracted from it.

In general, the velocity maps show a complex morphology in all channels, with many knots and extended filaments visible. However, in the low velocity channels, the observed [N II]$\lambda 6584$/H$\alpha$ ratios are low, and much more uniform across the field. In the high velocity channels we see a rich system of bright knots with  [N II]$\lambda 6584$/H$\alpha > 7-10$. In the eastern lobe, these knots are generally extended in the radial direction, while in the western lobe, they are more compact and brighter, especially in position P17.

In the eastern lobe, the radial velocities reach extreme values of -340\,km/s in positions P12a, 13, 13a and 14 in the eastern lobe, while radial velocities as high as +390\,km/s are seen in positions P17 and P18 in the western lobe. By reference to Figure \ref{fig1} we see that all of these positions lie along the central axis of the bipolar structure, and are essentially equidistant from the central PN core. This is suggestive of a single outburst event as the origin of these knots.
Correcting the observed radial velocities for the projection angle, we estimate that the space velocity of these knots reaches 550\,km/s in the east, and 650\,km/s in the western lobe.

The values of the [N II]$\lambda 6584$/H$\alpha$ in the halo reach similar values to those seen in the core (see Figure \ref{fig2}). This strongly suggests that the high velocity blobs and filaments have been ejected from this core in recent times. Assuming that the fastest knots have travelled at constant velocity out to the bipolar shell, then, using the angle subtended on the sky and assuming $i=38^o$, they have travelled 5pc since the outburst. At a mean speed of 600\,km/s, this implies the outburst  occurred $\sim 8100$\,yr ago.

This age is comparable to what might be expected for the post-AGB age of the PN, so we infer that the material with velocities close to systemic, and with much lower [N II]$\lambda 6584$/H$\alpha$ ratios, must represent swept-up gas left from an earlier ejection phase. However, this presents something of a paradox, since we might expect this swept-up gas to be also expanding at high velocity. We will return to this problem in the discussion.

This inferred age of $\sim 8000$\,yr clearly excludes the intriguing possibility discussed by \citet{Webster78} that the PN Hen 2-111 could be associated with the new star of AD\,185. If the ejected material had been following a Sedov-like expansion, rather than being in free expansion, the minimum age of the bipolar structure would be even greater,  $\sim 20000$\,yr.

\subsection{Excitation Analysis of the Bipolar Halo}

We have observed contiguous regions in each of the two lobes, P13, P13a and P12a in the eastern lobe, and P17, P18 and P19 in the western lobe. These regions encompass some of the highest-velocity gas. Particularly striking is N-rich jet-like structure which appears in the $v=-246$\,km/s channel in P12a. At low velocity we see long aligned filaments of [O III]-emitting gas which can be traced back into P13a and P13 -- see Figure \ref{fig8}. We interpret this as the result of a dense N-rich ``bullet'' passing through and shocking the ambient medium. Such a ``bullet'' would heat the interstellar gas up to $\sim 2.5\times 10^6$\,K, which would then expand laterally, driving a relatively slow cocoon shock in the interstellar gas, with a compressed shell of cooled gas behind it. This cooled, density-enhanced gas, photoionised by the central star of the PN could then become visible as the linear [O III] filaments seen in Figure \ref{fig8}.

\begin{figure}
  \includegraphics[scale=0.6]{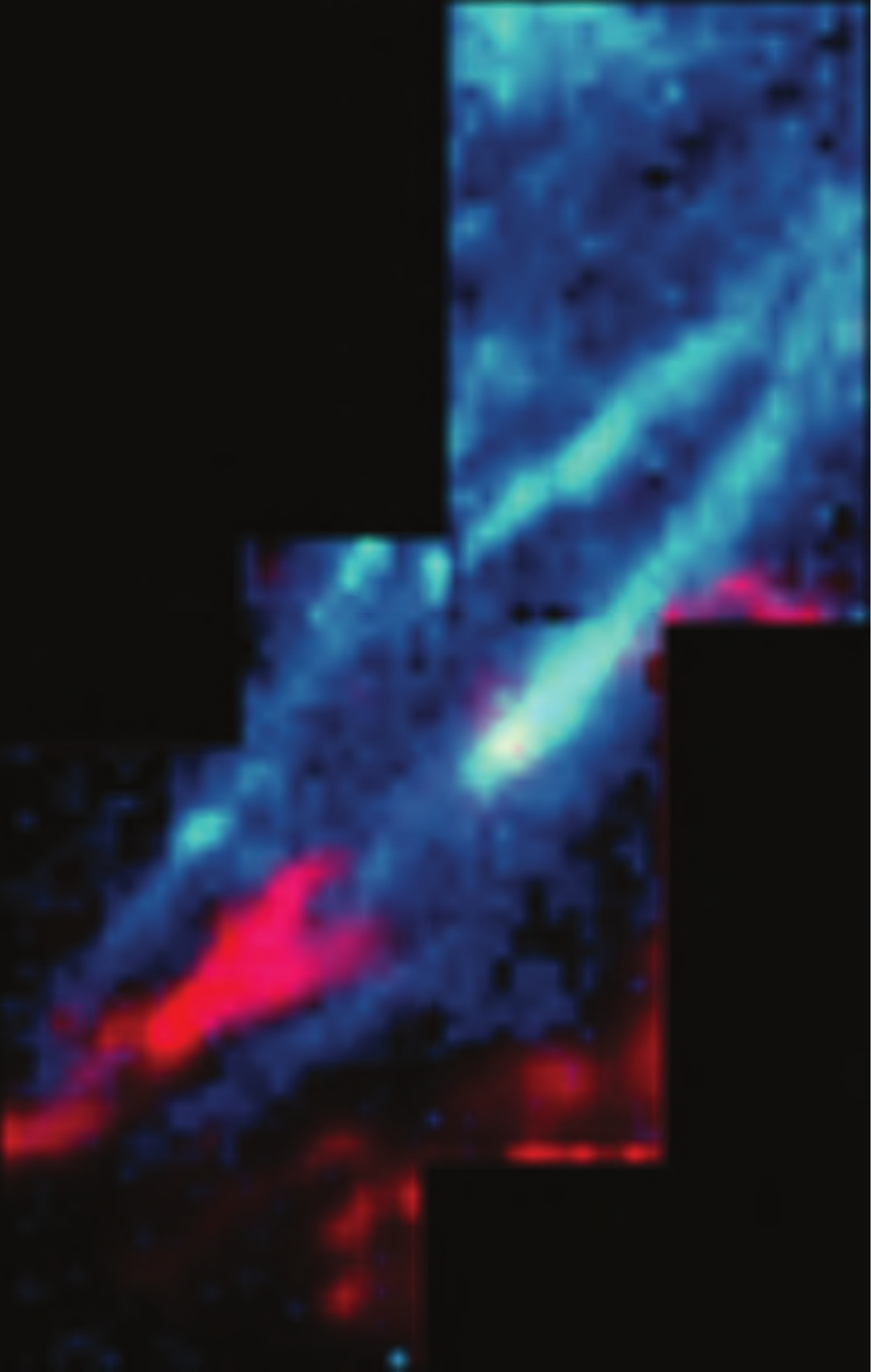}
  \caption{A colour mosaic of positions P12a, P13a and P13. The [N II] emission at a radial velocity of -246\,km/s (red) and the rest [O III] emission (green, blue). Here the fast moving knot of N-rich material is clearly exciting a radiative cocoon shock in the ambient interstellar material. This is discussed further in Section \ref{phase}.} \label{fig10}
 \end{figure}
 
 \begin{figure*}
  \includegraphics[scale=0.4]{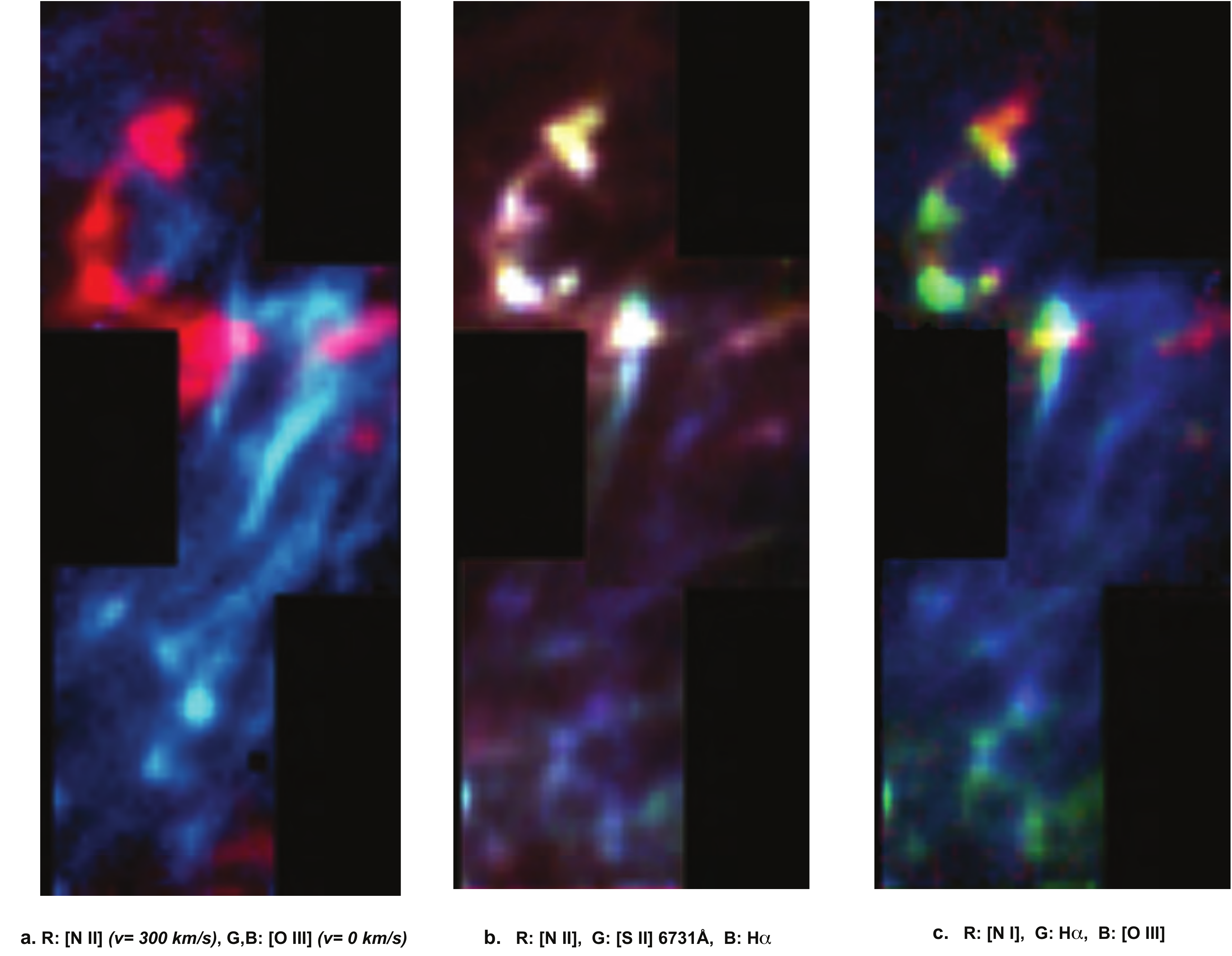}
  \caption{Colour mosaic of positions P17, P18, and P19 Left: The [N II] emission at a radial velocity of +300\,km/s (red) and the rest [O III] emission (green, blue). Centre:  Integrated [N II] emission (red), [S II]$\lambda6731$ emission (green) and H$\alpha$ (blue). Right: Integrated [N I]$\lambda\lambda 5198, 5200$ emission (red), H$\alpha$ (green) and [O III] emission (blue). Note that the N-rich fast moving material is concentrated in an arc, suggestive of an ongoing interaction with a dense interstellar cloud.} \label{fig11}
 \end{figure*}
 \begin{figure*}
  \includegraphics[scale=0.4]{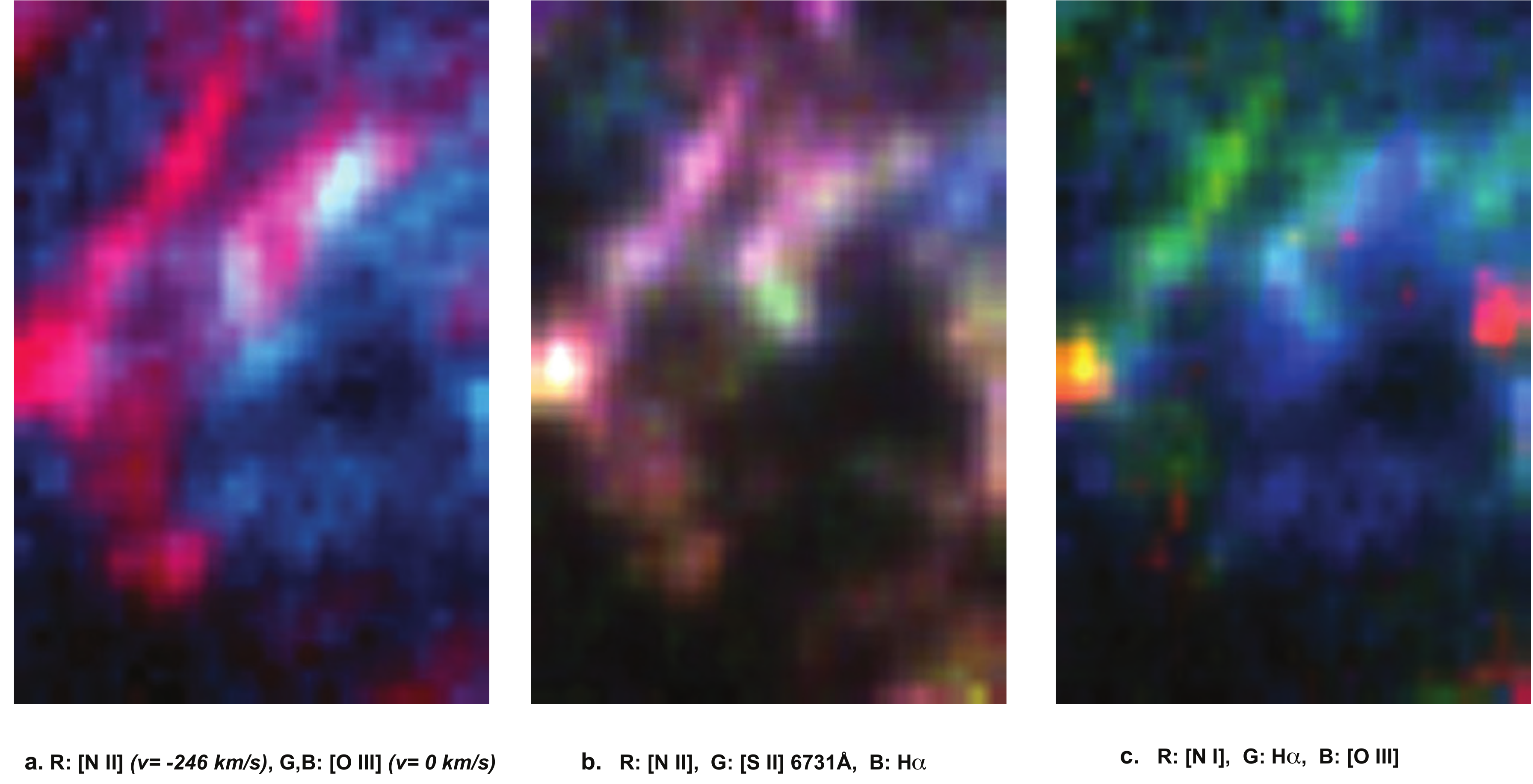}
  \caption{Colour images of position P12. Left: The [N II] emission at a radial velocity of -246\,km/s (red) and the rest [O III] emission (green, blue). Centre:  Integrated [N II] emission (red), [S II]$\lambda6731$ emission (green) and H$\alpha$ (blue). Right: Integrated [N I]$\lambda\lambda 5198, 5200$ emission (red), H$\alpha$ (green) and [O III] emission (blue).  Here the bright [N I]-emitting knot is close to the eastern edge of the field, while a fainter lower-velocity knot is visible at the western edge c.f. Figure \ref{fig3}..} \label{fig12}
 \end{figure*}

In the western lobe, the appearance of the knots is rather different. In Figure \ref{fig9} we present excitation maps of the high-velocity [N II] knots. Here the [O III] emission is essentially uncorrelated with the high velocity N-rich knots. These knots are distributed in an arc, and many of them also have strong [S II] emission. Most remarkably, these knots also display strong [N I] emission. 

Strong [N I] emission is only detected in one other position;  P12 in the eastern lobe. This is virtually in the diametrically opposite direction and at the same projected distance from the PN core as the knots in P17. The excitation images are given in Figure \ref{fig10}. This [N I]- bright knot has a trailing ridge of high velocity gas pointing back to the PN nuclear region.

In Figure \ref{fig11} we show the most interesting part of spectrum of the brightest northernmost knot in P17. The de-reddened line fluxes for this spectrum are listed in Table \ref{table2}. Here we have used the reddening function developed by \citet{Fischera05}, which is appropriate to a turbulent fractal foreground screen of dust. The derived logarithmic reddening constant $c$ is given at the foot of the Table. For this spectrum and for the other spectra extracted in this paper, the emission-line fluxes, their uncertainties, the velocity FWHMs and the continuum levels were measured using the interactive routines in {\tt Graf} \footnote{Graf is written by R. S. Sutherland and is available at: {\url {https://miocene.anu.edu.au/graf}} }and in {\tt Lines} \footnote{Lines is written by R. S. Sutherland and is available at: {\url {https://miocene.anu.edu.au/lines}}}. 

Note that in Figure \ref{fig11} the faint low-velocity component of emission is detected in the Hydrogen recombination lines, as well as in the [O III], [N II] and [S II] lines. However the high velocity component is much stronger, and this high-velocity component is the only one which is present in the [N I] doublet. The value of the [S II]$\lambda\lambda 6731/6717$ ratio is well above the low-density limit, and indicates an electron density  of  $n_e\mathrm{[S II]} \sim 600$\,cm$^{-3}$. Both low- and high-velocity components of the [O III] doublet are similar in intensity. We will attempt to model this knot in Section \ref{discussion}.

\begin{figure}
  \includegraphics[scale=0.42]{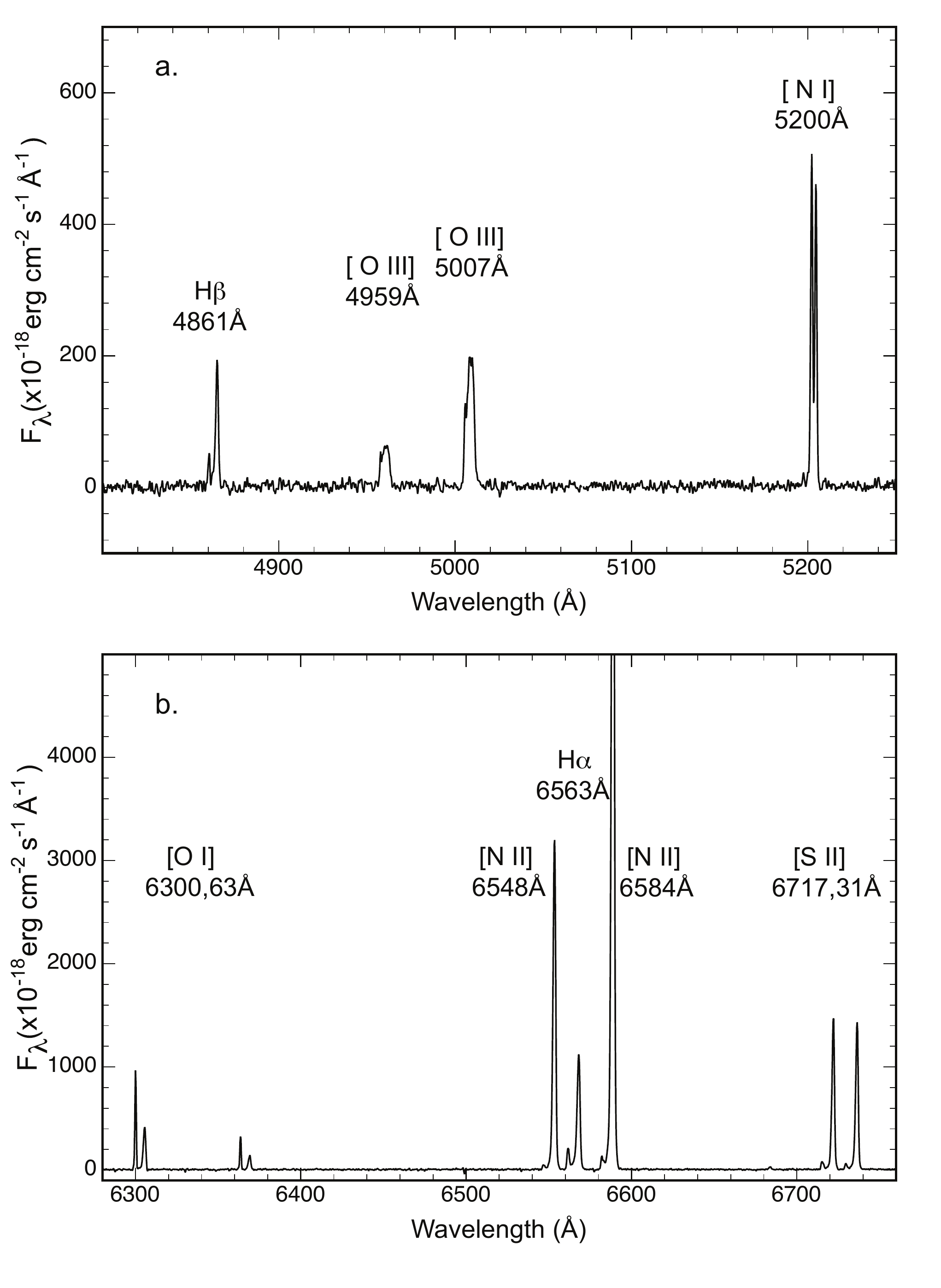}
  \caption{The spectrum of the northernmost knot in position P17, (a) in the blue and (b) in the red. Note that the [N I] doublet is much stronger than H$\beta$, and is associated with only the high velocity component of the line. Likewise the [S II] is stronger than H$\alpha$, and is also associated with the fast-moving material. The indicated electron density is  $n_e\mathrm{[S II]} \sim 600$\,cm$^{-3}$. The strong rest wavelength [O I] line is caused by residual night sky emission.} \label{fig13}
 \end{figure}

\section{Discussion} \label{discussion}
\subsection{Are the N-rich knots photoexcited?}\label{photo}
At first glance, one might suppose that the whole of the giant bipolar halo of Hen 2-111 is excited by the EUV photons coming from the central star. However, there are a number of surprising difficulties with this hypothesis, due to the weakness of the local EUV radiation field, as first indicated by \citet{Meaburn89}.

 Take for example the case of the complex of knots in positions P17 and P18, depicted in Figure \ref{fig7} and \ref{fig9}. These have a characteristic linear dimension of 5 arc sec., or about $1.5\times10^{17}$\,cm at the distance of Hen 2-111, and they lie some 5\,pc from the central star.

The position of the central star on the H-R Diagram has been determined by \citet{Marigo03} using the Helium Zanstra technique; $\log(L_*/L_{\odot}) = 2.66$ and $\log T_* = 5.24$\,K. (Here we have corrected the luminosity for  differences in the assumed distance to Hen 2-111 -- some 2.5\,kpc \citep{Marigo03} as compared with the 2.09\,kpc adopted here.

First, let us assume that the knots are optically thick, and construct a photoionisation model using the code MAPPINGS 5.12 \citep{Sutherland17}  \footnote{Available at : \newline {\url{https://miocene.anu.edu.au/mappings}}} initially adopting the abundance set given by \citet{Pottasch00} and \citet{Marigo03}. The constraint that the thickness of the ionised layer does not exceed the characteristic linear dimension determines the internal pressure ($\log P/k =5.9$\,cm$^{-3}$K) or, equivalently, the mean electron density ($n_{e\mathrm{[S II]}} =14$\,cm$^{-3}$) of the knots.

Such a photoionisation model completely fails to reproduce the observed properties of the knots. First, the ionisation parameter within the N-rich clouds is much too low to fully ionise hydrogen. A dusty photoionisation model has a peak HII fraction of only 0.43. As a consequence the low ionisation lines are much too strong relative to H$\beta$. For example, [N I]$\lambda \lambda 5198, 5200$/ H$\beta = 8.36$,  [O I]$\lambda 6300$/ H$\alpha = 0.78$ and  [S II]$\lambda \lambda 6717,31$/ H$\alpha = 4.9$, while [N II] is too weak,  [N II]$\lambda 6584$/ H$\alpha =3.46$. More seriously, the [S II]$\lambda \lambda 6717/6731$ ratio is at the low density limit, rather than $\sim 1.0$ as shown on Figure \ref{fig11}. This requires $n_{e\mathrm{[S II]}} \sim 600$\,cm$^{-3}$, rather than the 14\,cm$^{-3}$ implied by the model. Finally, the model fails to produce \emph{any} [O III] emission, which is again the consequence of the low ionisation parameter.

This unsatisfactory situation is alleviated only a little by assuming that the knots are optically thin to the EUV continuum. An extreme model with an optical depth of only 0.5 in the H-ionising continuum gives [N I]$\lambda \lambda 5198, 5200$/ H$\beta = 2.6$,  [O I]$\lambda 6300$/ H$\alpha = 0.36$, [S II]$\lambda \lambda 6717,31$/ H$\alpha = 2.4$, and  [N II]$\lambda 6584$/ H$\alpha = 4.06$. The issues both of the absence of [O III] and the very low electron density in the model remain.

\subsection{Reverse Shocks in the N-rich knots?}\label{Nshock}
The original hypothesis of \citet{Meaburn89} had fast N-rich bipolar ejecta interacting with denser knots. Our new dynamical evidence does not support such a  model, both because the fast moving material is concentrated in the N-rich
knots, rather than in the inter-knot medium, and also because the fast-moving knots display both higher surface brightness and density than the surrounding medium.

Given that some of the fast N-rich knots have a bullet-like appearance aligned with [O III] filamentary emission -- see Figure \ref{fig8} -- it is reasonable to suppose that they may be shock excited. In this model, the N-rich knots are denser than the surrounding medium, and excite a fast bow shock in their passage. Assuming a characteristic velocity of $\sim 600$\,km/s, the peak temperature reached in the bow-shock is of order $6\times10^6$\,K. This is a maximum temperature, since the bow shock may be at quite an oblique angle to the flow, as suggested by the [O III] morphology in Figure \ref{fig8}. 

The ram-pressure in this shocked gas drives a slower shock into the N-rich knot, which may become radiative during the interaction. It is this mechanism that can give rise to the excitation of the N-rich material, which we now investigate.
To investigate this we have built self-consistent radiative shocks with pre-ionisation using the MAPPINGS 5.12 code, the methodology described in \citet{Sutherland17}, and applied to the study of Herbig-Haro objects by \citet{Dopita17b}.

We model the spectrum of the bright knot in P17 shown in Figure \ref{fig10}, for which the the de-reddened line fluxes are listed in Table \ref{table2}.  The measured emission line width (FWHM = 2.2\,\AA) suggests that the shock velocity is in excess of 100\,km/s. In addition, the strong [O III] emission also implies that the shock velocity must be in excess of 100\,km/s. At these velocities Hydrogen is pre-ionised in the shock and the relative fluxes of the main optical emission lines vary only slowly with velocity \citep{Sutherland17,Dopita17b}.

\begin{table}
 \caption{The spectrum of the P17 knot, and the best-fit, finite-age shock models described in the text, all scaled to I(H$\beta$)=100.}\label{table2}
 \scalebox{0.8}{
\begin{tabular}{lclccc}
\hline
Lambda 	& Ion 	&Dered. & Shock & Shock & Shock\\
(\AA) 	&  	       &     Flux            & Model\#1$^{1}$ & Model\#2$^{2}$ & Model\#3$^{3}$ \\
\hline
4340.47	& H$\gamma$ & $40.3\pm 4.3$  & 46.0 & 45.9 & 45.6 \\
4363.21	& [O III] 	  & $20.4\pm 5.4$ & 11.1& 19.3 &  16.6   \\
4861.33	& H$\beta$ & $100\pm 4.5$  & 100 & 100 & 100 \\
4958.91	& [O III]	&  $55.4\pm 3.0$  & 38.0 & 69.1 & 60.7 \\
5006.84 	& [O III]     	&  $168\pm 8.6$  &  109.7 & 199.8 & 175.6 \\
5198,200 & [N I]	&   $181.1\pm 3.8$ &  72.6 & 67.5 &  67.0 \\
5754.59    & [N II]	& $63.1\pm 2.2$ & 21.8 & 33.7   & 45.2 \\
5875.66   & HeI 	&  $29.2\pm 2.1$ &  16.6 & 36.7 & 32.0 \\
6300.30   & [O I]	&  $85.1\pm 6.0$ &  15.8 & 15.5 &  6.0 \\
6363.78   & [O I]	&  $27.83\pm 4.9$ & 5.1 & 5.0  &  2.0 \\
6548.05    & [N II]	&  $700\pm 24$  &  367 & 537.1 &  576.0 \\
6562.82	& H$\alpha$ &  $283\pm 11$ & 301.4 & 303.7 &  307.3 \\
6583.45    & [N II]	&  $2217\pm 42$  &  1080 & 1582 &  1695 \\
6678.15    & HeI 	&  $9.4\pm 0.9$ & 4.7 & 10.5 &  9.15 \\
6716.44    & [S II]	&  $296\pm 7.0$ &  143.2 & 170.4 &  164.2 \\
6730.82    & [S II]	&  $295\pm 7.0$ &  143.0 &  189.6 &  161.4 \\
Red. Const.& $c=$ & $0.70\pm 0.1$ && & \\
& & & & & \\
\hline			
\end{tabular}} \\
$^{1}$ {$v_s = 150$km/s, $n(\mathrm H) =8$\,cm$^{-3}$, $\tau = 1000$\,yr} \\
$^{2}$ {$v_s = 150$km/s, $n(\mathrm H) =8$\,cm$^{-3}$, $\tau = 500$\,yr} \\
$^{3}$ {$v_s = 150$km/s, $n(\mathrm H) =6$\,cm$^{-3}$, $\tau = 630$\,yr} \\
\end{table}

We ran a grid of finite-age shocks with self-consistent pre-ionisation conditions, and using the \citet{Pottasch00} and \citet{Marigo03} abundance set. To ensure the correct pre-ionisation, the whole shock is bathed in the EUV radiation field of the central star. The shock velocity was kept fixed at $v_s =150$\,km/s, and the pre-shock density fixed at $n(H) = 8$\,cm$^{-3}$. The line intensities of the best-fit model is given as Shock Model \#1 in Table \ref{table2}. This model is of a partially-radiative shock of age 1000\,yr. 

While the model gives a reasonably good qualitative fit to the observed spectrum, \emph{all} the emission lines are predicted too weak relative to hydrogen in the model. Note in particular that the Helium recombination lines are also predicted too weak relative to hydrogen. The conclusion that can be drawn from this is that the N-rich knot in P17 contains partially hydrogen-burnt material. To investigate this, we ran grids with the hydrogen abundance by number reduced to 66\%, 50\%, 33\% and 25\% of its initial value. Because the higher chemical abundances of the heavy elements results in faster cooling, the shock age at a given final temperature scales roughly in proportion to the relative hydrogen abundance.

The best fit to the observations was found with the hydrogen abundance by number reduced to 50\% of its initial value, and for a shock age of 500\,yr. The relative fluxes resulting are given as Shock Model \#2 in Table \ref{table2}. All line fluxes lie much closer to the observed values, with the exception of the [O I] lines which are not well fitted. Importantly, the HeI fluxes relative to H are more nearly correct.

Finally, we ran a set of models in which we allowed the abundances of He, N, O and S to change, so as to better match the observations. These models were run at a slightly lower pre-shock hydrogen density ($n({\mathrm H}) = 6$\,cm$^{-3}$) so as to closely reproduce the observed [S II]$\lambda \lambda 6731/6717$ ratio. The best-fit model with a shock age of $\tau = 630\pm150$\,yr resulting from this exercise is given as Shock Model\#3 in Table \ref{table2}. This model has  hydrogen, helium and `metal' abundances of $X= 0.432$, $Y =0.542$ and $Z=2.61\times 10^{-2}$ respectively.  By number with respect to hydrogen, the abundances of the elements in the model are He: 0.316, N: $2.5 \times 10^{-3}$, O: $5.4 \times 10^{-4}$ and S: $5.0 \times 10^{-5}$. These abundances are consistent with both CN processing, and partial H-burning, which show that the material was derived from the innermost part of the H-burning shell.

From Table \ref{table2}, we note that \emph{all} shock models underestimate the strength of [O I] and to a lesser extent, that of [N I]. This is a generic problem with the current generation of shock models, caused by the fact that both of these species arise in a thin transition region of the shock in which the fractional ionisation and the electron temperature are decreasing rapidly. The line intensity predicted by the models therefore depends critically on the structure of this region. This issue could be alleviated in a number of ways, such as increasing the magnetic parameter, providing an additional source of heating in the transition zone, such as turbulent or conductive  heating, or simply by having a mixture of shock velocities, in particular, having shocks of lower velocity contributing to the emission \citep{Sutherland17}.

The minimum mass of the fast-moving cloud can be easily estimated. Adopting the above composition, density and computed age of the shock, we estimate that the shock has passed through a distance of $\sim3\times 10^{17}$\,cm and a column density of $\sim 7\times 10^{-6}$g\,cm$^{-2}$. Given that the transverse dimension of the cloud is $\sim 1.5\times10^{17}$\,cm, we estimate that its mass is $M_{\mathrm FM} \geqslant 8\times 10^{-5}$\,M$_{\odot}$. We cannot be certain as to the total number of high-velocity clouds, but we roughly estimate that there may be of order 20 of these in the whole bipolar shell. This would indicate a ejected mass $M_{\mathrm ej} \sim 0.002$\,M$_{\odot}$ or greater. This mass is appreciably greater than the  the mass predicted in the \citep{Soker92} BRET model; $\sim 10^{-4}~M_{\odot}$

\subsection{Excitation of the low-velocity [O III] filaments}\label{Oshock}
To understand the excitation of the low-velocity filaments which are bright in [O III] emission, we extracted the mean spectrum of the brightest of these from positions P18 and P19. The de-reddened spectra, and the logarithmic reddening constants used, $c$, are given in Table \ref{table3}. 

In P18 we used a circular extraction aperture of radius 5 arc sec centred on the pixel $x=14, y=21$ (measured from the E and N side of the field, respectively).The low-velocity [O III]-bright  filament so extracted is clearly visible in the central part of the image in Figure \ref{fig9}. This low-velocity feature is narrow; FWHM = 1.4\AA\ at H$\alpha$, corresponding to 65\,km/s, or $\sim47$\,km/s, correcting for the instrumental resolution.

In P19 we used a larger circular extraction aperture of radius 11 arc sec., centred on the pixel $x=13, y=22$ and encompassing the three brighter knots in the filamentary complex in Figure \ref{fig9}. At this position, the low-velocity feature is somewhat broader than in P18; FWHM = 1.95\AA\ at H$\alpha$, corresponding to 90\,km/s (78\,km/s correcting to the instrumental resolution). 

The [O III]-bright filaments have similar spatial widths to the N-rich knots, and so by the same arguments as in Section \ref{photo}, we conclude that they are unlikely to be photo-excited by the central star. Rather, it is more probable that here we are seeing a set of approximately edge-on cocoon shocks set up in the pre-existing material,  which have been excited by the passage of the high-velocity N-rich knots.

\begin{table}
 \caption{The (de-reddened) spectra of the P18 and P19 low-velocity filaments, and the best-fit, finite-age shock models described in the text, all scaled to I(H$\beta$)=100.}\label{table3}
 \scalebox{0.8}{
\begin{tabular}{lclclc}
\hline
Lambda 	& Ion 	&   P19 	& Shock &  P18 & Shock \\
(\AA) 	&  	         &    	          & Model\#4$^{1}$  & & Model\#5$^{2}$ \\
\hline
4340.47	& H$\gamma$ & $46.9\pm 9.1$ & 45.4 &  $45.3\pm 7.1$ &  43.1  \\
4363.21	& [O III] 	        & $33.5\pm 4.0$ & 34.3  & $41.6\pm 5.8$ &   45.0  \\
4861.33	& H$\beta$      & $100\pm 5$   & 100  & $100\pm 6$ & 100  \\
4958.91	& [O III]	        & $134.2\pm 4.0$    & 141 & $254\pm 9$ &  184 \\
5006.84 	& [O III]     	        & $408.8\pm 15 $  & 408 &  $796\pm 34$ &  533 \\
5754.59    & [N II]	        & $16.4\pm 2.3$    &  24.5 & $20.1\pm 4.4$ &    29.3 \\
5875.66   & HeI 	        & $12.2\pm 2.0$   &  17.3 &  $15.1\pm 3.3$ &  10.1 \\
6548.05    & [N II]	        & $146 \pm 5.1$ &  193 &  $139\pm 15$  &  180 \\
6562.82	& H$\alpha$   & $311 \pm 11$   &  300 & $306 \pm 11$ &  339 \\
6583.45    & [N II]	       & $468 \pm 17$ &  570 &  $529 \pm 19$& 529 \\
6716.44    & [S II]	       & $230\pm 18$ &  214 & $155 \pm 5$ &   168\\
6730.82    & [S II]	       & $174\pm  14$ &  154 &  $111\pm 4$  &   122 \\
Red. Const.& $c =$ & $0.90\pm 0.08$ && $0.8 \pm 0.05$ & \\
& & & & & \\
\hline			
\end{tabular}} \\
$^{1}$ {$v_s = 80$km/s, $n(\mathrm H) =4$\,cm$^{-3}$, $\tau = 1200$\,yr} \\
$^{2}$ {$v_s = 80$km/s, $n(\mathrm H) =4$\,cm$^{-3}$, $\tau = 600$\,yr} \\
\end{table}

We model the spectrum of P18, we assumed a shock age the same as we inferred for the shock in the fast moving knot of P17, $\tau =600$\,yr, and set a fixed shock velocity of 80\,km/s to be compatible with the observed velocity dispersion in both P18 and P19. We then investigate the sensitivity of the spectrum to changing pre-shock density. 

The P19 filaments are somewhat more radiative than the P18 filaments. This is evident both from the spectrum in Table \ref{table3} and from the excitation images of Figure \ref{fig9}. The [O III] lines are weaker with respect to H$\alpha$ or H$\beta$, and the [S II] lines are relatively stronger. Therefore, we assumed the pre-shock density to be the same as in P18, and varied the shock age. The best fit model gave $n(\mathrm H) =4$\,cm$^{-3}$ and $\tau = 1200$\,yr. However, an equally good fit could be obtained with a higher pre-shock density $n(\mathrm H) =6$\,cm$^{-3}$ and a smaller shock age $\tau = 600$\,yr, since both these shocks are at a similar phase of their  radiative evolution. The best-fitting models are given in Table \ref{table3}, as Shock Model \#4 (the fit to P19) and Shock Model \#5 (the fit to P18). 

The chemical abundances in all these shock models are assumed to be the same in both the P18 and P19 filaments, but are optimised during the fitting process. This gave hydrogen, helium and `metal' abundances of $X= 0.495$, $Y =0.495$ and $Z=1.0\times 10^{-2}$ respectively. These can be compared with the values derived for the fast moving N-rich P17 knot; $X= 0.432$, $Y =0.542$ and $Z=2.61\times 10^{-2}$ respectively.  Clearly, the medium containing the low-velocity [O III] filaments is less chemically evolved than the fast-moving filaments, but is clearly derived from the central PN, rather than from the general ISM.

By number with respect to hydrogen, the abundances of the elements in the model for the P18 and P19 filaments are He: 0.25, N: $1.7 \times 10^{-4}$, O: $3.1 \times 10^{-4}$ and S: $1.6 \times 10^{-5}$ (c.f. He: 0.316, N: $2.5 \times 10^{-3}$, O: $5.4 \times 10^{-4}$ and S: $5.0 \times 10^{-5}$ derived from the fast-moving N-knot in P17). These derived abundances are much closer to, but still systematically higher than, the \citet{Pottasch00} and \citet{Marigo03} abundance set derived for the core of Hen 2-111 -- He: 0.185, N: $3.0 \times 10^{-4}$, O: $2.7 \times 10^{-4}$ and S: $1.5 \times 10^{-5}$.

\subsection{The phase structure of the bipolar shell}\label{phase}
It is interesting that the ram-pressure inferred for the shocks in the fast-moving material ($P_{FM} \sim 4\times 10^{-9}$\,dynes/ cm$^2$) is only a little higher than that inferred for the ``rest" [O III] filaments ($P_{Fil} \sim 0.9\times 10^{-9}$\,dynes/ cm$^2$). Given that the  fast-moving N-rich knots have a space velocity of $\sim 600$\,km/s, they must be producing a bow shock with a velocity of this order. Equating the ram pressure of cloud shock to the ram pressure at the leading edge of the bow shock, this would imply a pre-shock density in the surrounding medium of only $n(\mathrm H) \sim0.36$\,cm$^{-3}$, which is an order of magnitude lower than inferred for the low velocity [O III] filaments. We must therefore conclude that the interstellar medium (ISM) of the lobes of Hen 2-111 is a two-phase medium, and that the low-velocity filaments are shocks in the denser phase which have been excited by thermal pressure in the low-density phase resulting from bow-shocks caused by the passage of the high-velocity knots.

Such shocks may arise in the form of cocoon shocks driven by the hot plasma generated by the fast shock in the low density phase of the ISM. The post-shock temperature at the leading edge of the bow shock may be as high as $6\times10^6$\,K. This gas flows back from the high velocity cloud, and expands laterally to form a cocoon shock. The theory of this  type of flow is familiar in the context of the evolution of radio lobes in active galaxies \citep{Begelman96, Bicknell97}. The mean pressure driving the cocoon shocks is lower than the ram pressure at the leading edge of the flow by a factor $\zeta$ where typically  $\zeta \sim$ 2 - 10. This factor is very similar to the ratio derived here for the ram pressure in the N-rich cloud shock to the ram pressure inferred for the shock in the [O III] filaments; $\sim 5$.

The clearest example of this type of geometry is seen in Figure \ref{fig8}. Here the fast-moving N-rich knot in position P12a is surrounded by a cocoon with a projected half-cone angle of $\sim 10^o$. Correcting for the adopted inclination of the system $i =38^o$ given in Section \ref{chemo}, the true half-cone angle is  $\sim 8^o$. If this cocoon represents the partially-radiative shocks in an underlying medium, then in the co-moving frame of the fast N-rich knot (600\,km/s), the [O III] filaments represent oblique shocks with a velocity of $\sim85$\,km/s. This agrees well with the velocity inferred from the [O III] line widths for the filaments in P19, and with the shock velocity adopted in the models of the filaments in both P18 and P19. 

It may well be that in this case a low density ISM phase is not required, and that the partially-radiative cloud shock in the N-rich material is also oblique. Being of similar pre-shock density both the bow shock and the cloud shock at the leading edge of the N-rich cloud would then be non-radiative, giving rise to the hot gas which fills the cocoon.

\begin{figure}
  \includegraphics[scale=0.33]{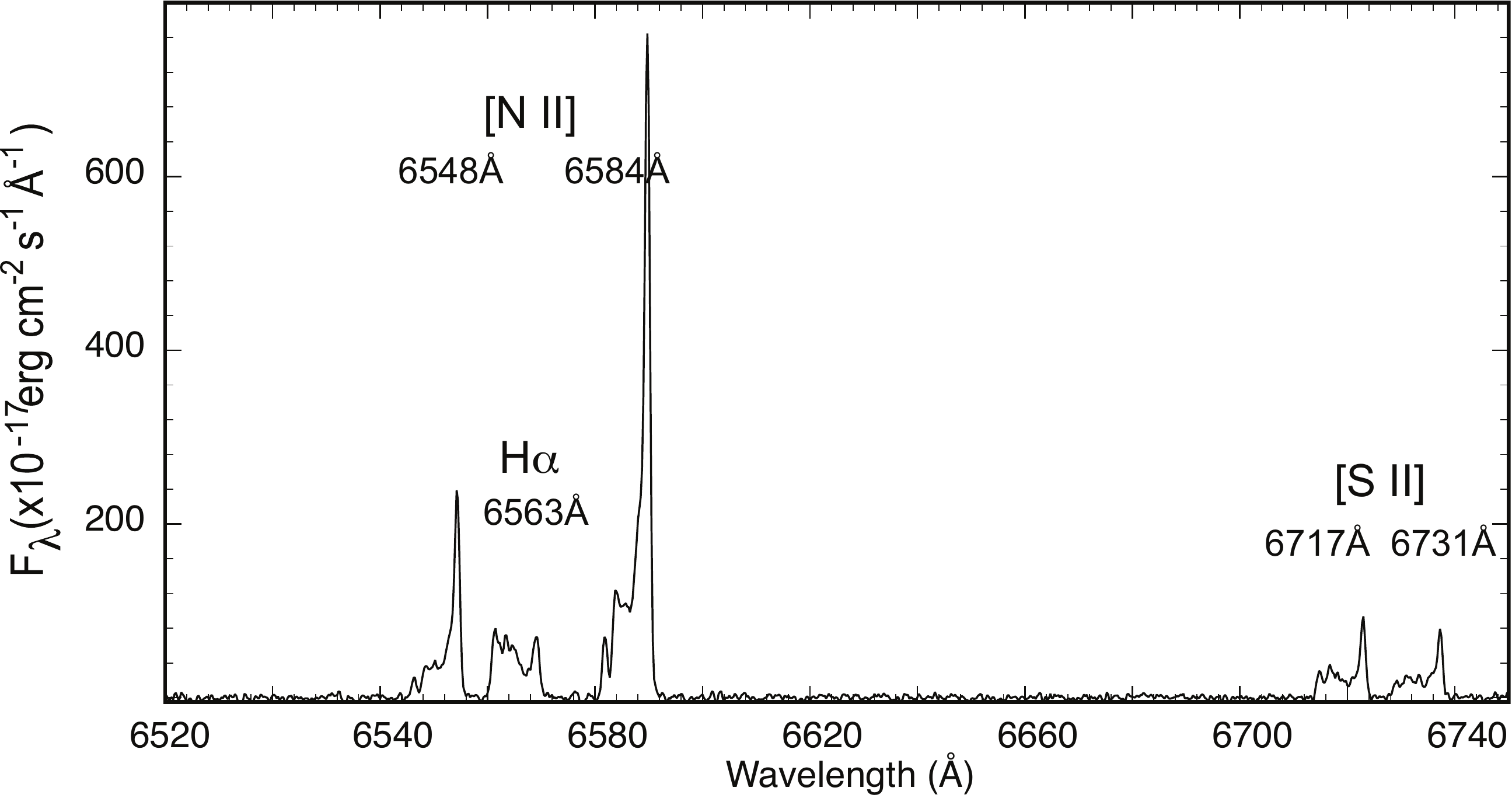}
  \caption{A portion of the red spectrum from the region to the SE of the bright knot in position P18. Note that the [N I] doublet is much stronger than H$\beta$, and is associated with only the high velocity component of the line. Likewise the [S II] is stronger than H$\alpha$, and is also associated with the fast-moving material. The indicated electron density is  $n_e\mathrm{[S II]} \sim 600$\,cm$^{-3}$. The strong rest wavelength [O I] line is caused by residual night sky emission.} \label{fig14}
 \end{figure}

Finally, there is some dynamic evidence that the fast N-rich knots are being ablated by their passage through the surrounding medium. The bright knot in P18 was earlier observed by \citet{Meaburn89}, who found continuous emission extending back to rest velocities (see their Fig 5g). In our data, we also find such a component. This is not visible in Figure \ref{fig9} due to the limitations of the scaling. The bridging emission is most prominent following the bright knot in the region located at the eastern extremity of the observed field (see Figure \ref{fig4} or Figure \ref{fig11}). We have extracted the spectrum of a 4 arcsec. circular aperture centred on this region, and show a portion of the red spectrum in Figure \ref{fig14}. Note that here the low-velocity H$\alpha$ is stronger than the high-velocity part of the line, while the reverse is true of the [N II]. The [S II] profiles are intermediate between these two extremes. The simplest  explanation of these profiles is that here we are seeing a turbulent ablative wake of the N-rich P18 knot, produced by the interaction of the knot with the surrounding material.

\subsection{Conclusions}
To summarise our findings in Sections \ref{results} and \ref{discussion}:
\begin{itemize}
\item{The giant bipolar lobes of Hen 2-111 displays strong spatial and dynamical related chemical fractionation as measured by the [N II]$\lambda6584$/H$\alpha$ ratio. All high velocity material shows strong enhancements in this ratio compared with the slow-moving material, and observed radial velocities of this gas decrease strongly away from the tips of the bipolar structure.}
\item{The fast-moving material achieves a maximum velocity $\sim 600$\,km/s, and is seen at a projection angle of $\sim 38^o$ to the plane of the sky, and was ejected from the core about 8000 years ago.}
\item{Most of the material in the bipolar lobes is too dense and too far from the central PN to be fully photoionised by the central star. The ionisation parameter in this gas is $U \sim 10^{-5}$, and hydrogen can be only partly ionised by the EUV photon field.}
\item{ The observed velocity dispersions and spectral properties of both the fast-moving N-rich knots and the low-velocity [O III]-bright filaments can be explained in terms of partially-radiative finite age shocks. The fast-moving knots are excited by a reverse shock moving into them, while the low velocity filaments are produced by the bow and cocoon shocks generated by the passage of the fast-moving N-rich ``bullets'' through the pre-existing bipolar halo of the PN.}
\item{The most rapidly moving knot in the polar direction is composed of partially hydrogen-burnt material with $X= 0.432$ and $Y =0.542$. This is passing through the bipolar halo gas which also partially hydrogen-burnt, although to not such an extreme degree ($X= 0.495$, $Y =0.495$). Both of these are more extreme than the central PN abundances inferred by \citet{Pottasch00} and \citet{Marigo03} ($X= 0.571$, $Y= 0.419$). Clearly therefore, the material ejected in the poleward direction originates from deeper within the hydrogen burning shell.}
\end{itemize}

It is clear that in Hen 2-111 we are seeing the effect of a violent final mass ejection event, on the cause of which we speculate here. The composition of the bipolar nebula halo also only shows the products of hydrogen burning nucleosynthesis, with a strong He and N enrichment and depletions in C and O relative to solar \citep{Marigo03}. Given that the central star of Hen 2-111 has an effective temperature and luminosity consistent with a core mass of $\approx 0.68-0.75 M_{\odot}$, this would indicate an initial progenitor mass of $\approx 3M_{\odot}$. However, single AGB stellar evolution theory predicts that such an object should be carbon rich, and would not show such strong He and N overabundances, relative to solar -- see the review by  \citet{Karakas14}.

We speculate here that binary evolution has  truncated evolution on the thermally-pulsing AGB before any third dredge-up could occur. The most likely explanation for the extreme velocities in the poleward direction and the chemical fractionation we see in the giant bipolar halo of the nebula are jets, which could have been launched during the common envelope phase  such as in the bipolar rotating episodic jets model of \citet{Soker92}. Alternatively, the common envelope phase could have led to an accretion disc forming around the white dwarf, which could easily drive outflows of 600 km/s.

\citet{Podsiadlowski10} suggest a mechanism of explosive common-envelope ejection in which a low-mass companion directly transfers hydrogen-rich matter during the common envelope phase onto the core of the primary star, of which is $0.03-0.06M_{\odot}$ is then explosively ejected. In principle, a similar situation could also occur in the case of an AGB binary. If the hydrogen enters the He-burning shell, perhaps as a result of a late He-shell flash, then this could result in an episode of explosive H-burning. The chemical products of this process, mixed with partial He-burning could then explain both the chemical abundances and the velocities of the N-rich knots.

The composition of the bipolar halo is still unexplained. The initial mass of the progenitor seems consistent with a mass closer to $\approx 3 M_{\odot}$ which is too low for hot bottom burning  \citep{Karakas14}.  However, given the uncertainty in the assumed distance, a consequent small shift in the estimated luminosity would place the post-AGB star on an evolutionary track consistent with a more massive progenitor. This may then allow for thermal pulses and hot bottom burning to begin before the star went through the common envelope. A more precise estimate of the luminosity and effective temperature is required in order to verify this possibility.

Finally, \citet{Bear17} have examined the morphology of a number of Galactic planetary nebulae including Hen 2-111 and conclude that this object likely formed in a triple system, given the asymmetries in the nebula. This is consistent with the recent study by \citet{Garcia-Segura16} who show that spin-up in a binary system alone cannot explain the bipolar shape of PN. 
 
\section*{Acknowledgements}
The authors thank Christopher Tout and Sung-Chul Yoon for informative discussions on the theory of binary evolution.This research has made extensive use of NASA's Astrophysics Data System (ADS). 
\end{document}